\documentclass[aps,twocolumn,showpacs,amsmath,amssymb,floatfix,superscriptaddress]{revtex4}

\usepackage{graphicx}
\usepackage{dcolumn}
\usepackage{bm}
\usepackage{latexsym}
\usepackage{latexsym,graphics}

\newcommand{\be}{\begin{equation}}
\newcommand{\ee}{\end{equation}}

\begin{document}

\title{Communication and Synchronization in Disconnected Networks with  Dynamic Topology: Moving Neighborhood Networks}

\author{Joseph D. Skufca}
\email{skufca@usna.edu}
\affiliation{Department of Mathematics, United States Naval Academy, Annapolis, Maryland}%

\author{Erik M. Bollt}
 \homepage{http://www.clarkson.edu/\~bolltem}
\email{bolltem@clarkson.edu}
\affiliation{Departments of Mathematics \& Computer Science, and Physics, Clarkson University, Potsdam, NY 13699-5815}%

\date{\today}

\begin{abstract}

We consider systems that are well modelled as a networks that evolve in time, which we call {\it Moving Neighborhood Networks}.  These models are relevant in studying cooperative behavior of swarms and other phenomena where emergent interactions arise from ad hoc networks. In a natural way, the time-averaged degree distribution gives rise to a scale-free network. Simulations show that although the network may have many noncommunicating components, the time-averaged communication is sufficient to yield robust synchronization of chaotic oscillators.    
\end{abstract} 

\maketitle

Network dynamics has become a very important area in nonlinear studies because so many systems of interest have a natural description as a network.  Examples include the internet, power grids, neural networks (both biological and other), social interactions, and many more.  However, the preponderance of the work in complex networks does not allow for dynamic network topology \cite{NewmanReview,BarabasiReview}. To our knowledge, there has not been published any significant work on time-dependent network architecture, and none which considers a natural model of meandering agents.   In the literature, one finds that either a static network is `born', as in the small-world  (SW) \cite{Watts}  and Erdos-Renyi \cite{erdosrenyi} models, or that a network evolves into an otherwise static configuration, as is assumed in the Barabasi-Albert model of scale-free (SF) networks evolution \cite{barabasi1}.   In epidemic modeling and percolation theory, one considers the problem of certain links being knocked-out, but essentially as a static problem, since there is no possibility for links to reform within the theory.   

In this letter we consider a simple model that allows network links to follow their own dynamical evolution rules, which we consider a natural feature of many natural and technological networks, where autonomous agents meander or diffuse, and communication between them is an issue of both geography and persistence.  For our concept problem, we focus on social interactions, such as a disease propagating across a network of social contacts.  A suitable model should consider disease life-cycle, which may be just a matter of a few weeks; an epidemic ensues only if agents connect within that time window.  Thus, with expiring messages, what matters is whom we have {\it recently} contacted.  No current network models appropriately account for this realistic feature.   When the time scale for  network changes is of the nearly the same order as the time scales of the underlying system dynamics, we believe that network evolution should be part of the model.  We will study such networks using a model which we first introduce here and which we call {\it moving neighborhood networks} (MN).  We then modify the basic MN structure by considering that some social connections survive even when the local neighborhood has changed, which we call the {\it moving neighborhood with friends} model (MNF).   

We evolve (diffuse) positions of agents independently according to a dynamical system, or stochastic process, linking those nodes that are within the same neighborhood.  We assume that the system has an ergodic invariant measure, then we prove that the relative positions of the nodes, and hence their connectivity, is essentially random, but with a well defined {\it time average degree distribution.} Using the property of synchronization as a probe of connectivity, we show a most striking feature of such networks is that while at any fixed time, the network may be fractured into noncommunicating subcomponents, the evolving network allows communication of those messages which do not expire on time-scales at which a message can find paths between subcomponents; this point is made clear by the fact that MN and MNF networks admit surprisingly robust synchronization well below the threshold when the network has a giant component.

We consider two distinct dynamical systems: 1) the dynamical system which governs the network topology by diffusing agents corresponding as the indexed vertices, which we call {\it network dynamics,} and 2) the network of the oscillators which run at each vertex, with coupling between them moderated by the instantaneous network configuration, which we call the {\it system dynamics.} The formalism of {\it master stability function} \cite{Fink} (which assumes a fixed network) must be modified to consider evolving networks. Synchronization requires sufficient information flow, so complete paths must appear on time scales relevant to the {\it system dynamics}. We introduce the concept of a {\it moving average Laplacian} to quantify the connectivity associated with a time sensitive message that propagates on a partially connected but changing network.

{\bf The Moving Neighborhood Network:}\label{sMNdescription} Our modeling goal is to capture some features of evolving social networks.  Using an analogy of collaborators: often we work as part of a group at our place of employment, usually dealing with a small group of people (our neighborhood). People move from one job to another, so our neighborhood will frequently change, but only a little.  However, if {\it we} move, our neighborhood changes significantly; we end up with a completely different group of coworkers.  The time scale of the small changes is of the same order as might be required to tackle significant problems and therefore are relevant to the overall productivity of the group. 

To capture this dynamic, we associate network nodes with an ensemble of points evolving under a flow, forming a time dependent network by linking nodes that are `close.' Notationally, let $\bar{ \xi}=\{ \xi_1, \ldots, \xi_n \}$ be a collection of $n$ points in some metric space ${\cal{M}}$.  We construct graph ${\Xi}$ from $\bar{\xi}$ by associating a vertex with each element of $\bar{\xi}.$  Vertices $i$ and $j$ of $\Xi$ are assigned to be adjacent (connected by an edge) if $|\xi_i-\xi_j|<r,$ where $r$ is a parameter that defines the size of a neighborhood.  Let $\phi_t : {\cal{M}} \mapsto {\cal{M}}$ be the flow of some dynamical system on ${\cal{M}}.$   From an initial ensemble $\bar{\xi}^0=\{ \xi_1^0, \ldots, \xi_n^0 \},$ we define an ensemble trajectory by 
\begin{equation}
\bar{\xi}(t)=\Phi_t(\bar{\xi}^0)= \{ \phi_t(\xi_1^0), \ldots,\phi_t( \xi_n^0 )\},
\end{equation}
which in turn generates a graph $\Xi(t)$ that precribes a {\it network trajectory.}  The flow $\phi_t$ may be governed by any discrete or continuous time diffusive process, either deterministic or stochastic.  For brevity of presentation, we describe the MN process by using a deterministic, ergodic map ($\gamma$), representative of strobing a continuous diffusive system. Under suitable choice of $\gamma,$ most ensembles will distribute according to some natural invariant density, $\rho_\gamma,$ giving a well defined time-average network character. 

\begin{figure}[htbp]
\includegraphics[width=0.5\textwidth]{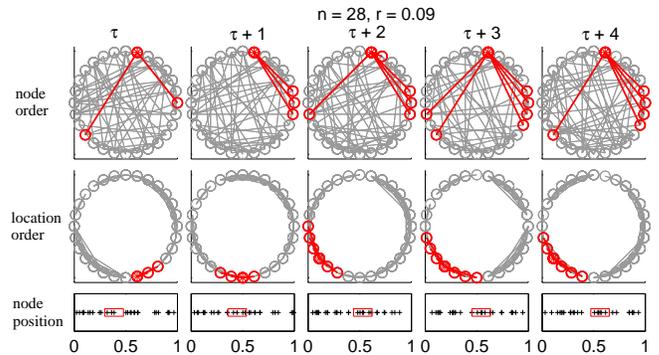} 
\caption{(Color online) An MN simulation using $n=28,$ $r=.09,$ and map of (\ref{egamma}).  Five time steps are shown.  The 1st row shows the network in node index order.  The 2nd row is the same network, with the nodes positioned by $\xi_i$ ordered to illustrate that they are neighborhood graphs.  The (bold) red portion of the network shows connections to node 1.  The bottom row shows the ensemble distribution of $\bar{\xi}.$  The relatively small $n=28$ was chosen for artistic reasons (to more easily display the connections).}
\label{fnetseq}
\end{figure}

{\bf Simulation: a specific MN network.} Consider the following construction:  Let $M={\rm T}^1,$ the circle, and let
\begin{equation}
\label{egamma}
\gamma(x):=1.43x-.43\frac{\lfloor 4x \rfloor}{4} \bmod 1.
\end{equation}
This map, chosen primarily for illustrative reasons, has the following characteristics: (1) it is choatic, (2) transitive on the invariant set $[0,1]$, (3) uniformly expanding, (5) with non-uniform invariant density, and (5) is discontinuous (so that a node may be moved to a distant neighborhood on one iteration).  From a random initial condition for $\bar{\xi},$ we iterate past the transient phase so that the ensemble resembles the invariant density.  We then construct the associated network for each iteration of map $\gamma.$ Fig \ref{fnetseq} shows the network constructed from five successive iterations, using $n=28$ and $r=0.09.$ Note that from one iteration to the next, the connections associated with node 1 change very little.  The reindexing and redrawing in the second row makes clear that the network is a neighborhood graph, though not all neighborhoods contain the same number of nodes.  Note that all but the $\tau + 2,$ the graphs have a disconnected component. 

{\bf Synchronization of Coupled Oscillators:} To explore the implications of an MN structure, we use synchronicity as a connectivity persistence probe of a network of $n$ identical chaotic oscillators.  We form a time dependant network, described by graph $G(t)$, consisting of $n$ vertices $\{v_i\}$, together with the set of ordered pairs of vertices $\{(v_i,v_j)\}$ which defines the edges.  The $n\times n$ adjacency matrix defines the edges, $A_{i,j}(t)=1$ if there is an edge $(v_i,v_j)$ at time $t$, and $=0$ otherwise.  The system of $n$ oscillators is linearly coupled by the network as follows: Let the vector ${\bf x}_i$ be the state vector for the $i$th oscillator and express the coupled system as
\begin{equation}
\dot{{\bf x}}_i(t)=f({\bf x}_i(t))+ \sigma \sum_{j=1}^n {L}_{ij}(t)K{\bf x}_j(t),
\end{equation}
where $\sigma$ a control parameter, $L_{ij}(t)$ the element of the graph Laplacian, $L(t)=Diag(d)-A(t),$ and $K$ specifies which state vector components are actually coupled.   Thus we have a dynamical system flowing on $\Re^{3n}$.  Specifically, we consider the R\"{o}ssler attractor with $a=0.165, b=0.2,c=10.0,$ which exhibits a chaotic attractor with one positive Lyapunov exponent \cite{Rossler}.  Coupling the $n$ systems through the $x_i$ variables, the resultant system is given by
\begin{equation}
\begin{array}{l}
\dot{x_i}=-y_i-z_i-\sigma \sum_{j=1}^n L_{ij}(t) x_j \\
\dot{y_i}=x_i+ay_i \\
\dot{z_i}=b+z_i(x_i-c). 
\end{array}
\end{equation}
Then the question of whether the oscillators will synchronized is reduced to whether one can find a value for $\sigma$ such that the synchronization manifold is stable.

{\bf Known results for static networks.}  For a fixed network, necessary conditions for synchronization are well described by the approach in \cite{Barahona,Fink}, summarized as follows: The graph Laplacian matrix $L$ has $n$ eigenvalues, which we order as $0=\theta_0 \leq \ldots \leq \theta_{n-1}=\theta_{max}.$  Using linear perturbation analysis, the stability question reduces to a constraint upon the eigenvalues of Laplacian:  
\begin{equation}
\sigma \theta_i \in (\alpha_1,\alpha_2) \quad \forall i=1, \ldots , n-1,
\label{estaba}
\end{equation}
where $\alpha_1,\alpha_2$ are given by the {\it master stability function,} a property of the oscillator equations.  For $\sigma$ small, synchronization is unstable if $\sigma \theta_1 < \alpha_1;$ as $\sigma$ is increased, instability arises when $\sigma \theta_{max} > \alpha_2.$   By algebraic manipulation of (\ref{estaba}), one can show that if $\frac{\theta_{max}}{\theta_1} < \frac{\alpha_2}{\alpha_1}=:\beta,$ then there is some coupling parameter, $\sigma_s,$ that will stabilize the synchronized state.  For some networks, no value of $\sigma$ satisfies (\ref{estaba}).  In particular, since the multiplicity of the zero eigenvalue defines the number of completely reducible subcomponents, if $\theta_1=0,$ the network is not connected, and synchronization is not stable.  However, even when $\theta_1>0,$ if the spread of eigenvalues is too great, then synchronization may still not be achievable.  

{\bf Refinement to evolving networks.} Numerical simulations of the MN model indicate that synchronization can occur even when the network fails criteria of (\ref{estaba}) {\it at every instant in time.}   Apparently, the temporal mixing creates an average connectedness that allows the network to support synchronization.  

We conjecture that an appropriate quantification of the average connectiviy is given by the {\bf Moving Average Laplacian,} which we introduce here and define as the solution to the matrix initial value problem,
\be
\dot{C}(t)= L(t)-C(t), \quad C(0)=L(0).
\label{emalde}
\ee
We solve (\ref{emalde}) to write 
\be
C(t)=e^{-t}\left( C(0)+\int_0^t e^\tau L(\tau) d \tau \right).
\ee
We estimate $\lambda_1^*=E[\lambda_1(C(t))]$ and   $\lambda_{max}^*=E[\lambda_{max}(C(t))],$ and then use $\lambda_1^*$ and $\lambda_{max}^*$ with (\ref{estaba}) to determine an appropriate choice for $\sigma$ to achieve stable synchronization on a particular MN network, which we now find predicts well the synchronization according to the now modified master stability formalism.

\begin{figure}[htbp]
\includegraphics[width=0.5\textwidth]{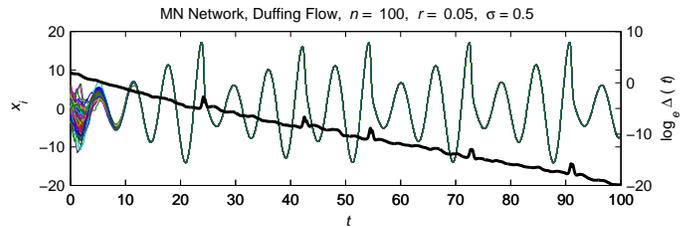}
\caption{(Color online) MN Network with $n=100$ nodes, $r=.05,$ with agents wander according to the chaotic Duffing equation, $x''=x-x^3-.02x'+3\sin t.$ The $x$-coordinate of each oscillator is plotted vs. time. The bold line is $\Delta,$ providing an estimated deviation from the synchronization manifold.}
\label{fduffT}
\end{figure}

{\bf Numerical explorations of MN behavior:}\label{sims}
Consider a system of $n=100$ agents wandering on the chaotic attractor of the Duffing equation, $x''=x-x^3-.02x'+3\sin t,$ whose driven frequency is commensurate with the natural frequency of the R\"{o}ssler system, $\omega \approx 1.$  If we choose $r=.05,$ the network is disjoint -- generally it has more than 25 disconnected components.  With the R\"{o}ssler systems starting from a random initial condition, Fig \ref{fduffT} shows a plot of $x_i(t),$ for the coupled system, {\it which indicates that despite the weak instantaneous connectivity of the network, the average effect results in synchronization.}  The bold curve illustrates the asymptotic stability of the synchronization state by graphing
\[
\Delta (t) = \frac{1}{n} \sum_{i=1}^n |x_i(t)-\bar{x}(t)|+|y_i(t)-\bar{y}(t)|+|z_i(t)-\bar{z}(t)|,
\]
where $(\bar{x}(t),\bar{y}(t),\bar{z}(t))=\frac{1}{n} \sum_{i=1}^n (x_i(t),y_i(t),z_i(t)) $ estimates the synchronization manifold. As an additional statistic, we define $k_{eq},$ to be $1/2$ the number of connections per node, averaged over space and time, which allows direct comparison with the value $k$ of a regular graph. We calculate $k_{eq} \approx 1.1$ (in comparison to a regular network, which would require $k \geq 7$ to synchronize); on average, a node is coupled to only a few other nodes on each time step.  However, the rapidly changing laplacian compensates.  Results are similar for other ergodic systems used to control agent flow, such as $\gamma(x)$ in Eq.~(\ref{egamma}).  

{\bf Analysis and conjectures:}\label{conj}
A counterintuitive result from the simulations is that although the synchronized state may be linearly unstable, the MN network can still synchronize.   The instantaneous interpretation is that an ensemble of conditions near the manifold is expanding in at least one direction, but is generally contracting in many other directions.  When the network reconfigures, the expanding and contracting directions change, so points in the ensemble that were being pushed away at one instant may be contracted a short time later.  If there is sufficient volume contraction and change in orientation of the stable and unstable subspaces, the MN network can achieve asymptotic stability despite linearly instability.  In the following paragraph, we give some mathematical basis of the above by considering a simple linear system description of variations from the synchronization manifold.

Consider the $n$ dimensional initial value problem
\begin{equation}
\dot{z}=A(t)z, \qquad z(0)=z_0,
\label{esys1}
\end{equation}
where $A(t)=\sum_i \chi_{[iT,(i+1)T]}(t)A_{i}$ is a piecewise constant matrix, $i$ an integer, and $T$ constant.  For narrative simplicity here, assume $A_i$ is a diagonal matrix, $A_i= diag\{\lambda_{i1},\ldots,\lambda_{in}\}.$ Since diagonal matrices commute, we may write the time $t_k=Tk$ solution to (\ref{esys1}) as
\begin{equation}
z(t_k)=e^{\int_0^{t_k} A(\tau)d\tau} z_0=e^{(A_0+ \cdots + A_{k-1})T} z_0.
\end{equation}
The fundamental solution matrix is diagonal with entries $\lambda_j = e^{s_{jk}},$ with $s_{jk}=\sum_{i=0}^{k-1} \lambda_{ij}T,$ and each $j$ can be associated with a coordinate direction in ${\rm R}^n.$ Stability of the origin is ensured if $s_{jk}$ is bounded above for all $j$ and $k.$  If $s_{jk} \to -\infty,$ then the origin is asymptotically stable.  Suppose the $A_i$'s are chosen ergodically from a distribution such that for all $i,$ $tr(A_i) < \epsilon <0.$  Moreover, assume that the positive and negative eigenvalues are distributed ergodically along the diagonal elements of $A_i.$  Then the time average (over $i$) must be the same as the spacial average (over $j$) of the eigenvalues, which implies that $s_{jk} \to - \infty.$  Since $det(\Phi(t_2,t_1))=e^{\int_0^t tr(A(\tau)) d\tau} < 1,$ we have that the system is volume contracting.

{\bf Time-Average Scale-Free Network.}  
The main thrust of this modeling effort is to show that it is useful to consider evolving networks.  The  underlying time average degree distribution  remains very flexible, including possibility of the scale-free distribution seen so frequently in many applications, \cite{NewmanReview,BarabasiReview}.  
The basic MN network generates a binomial degree distribution, seen easily as follows.  The probability that node $j$ is $\epsilon>0$ close to node $i$, which is at a position $x$ is, $p(x,\epsilon)=\int_{x-\epsilon}^{x+\epsilon} d\mu(x),$ (by assuming the network has the ergodic invariant measure $\mu(x)$).  The `long-run' probability that $i$ and $j$ coincide to within $\epsilon$ is
$p(\epsilon)=\int p(x,\epsilon) d\mu(x),$ where $p\equiv p(\epsilon)$ is simply some function of $\epsilon$. Therefore, the time-average degree distribution of MN is the binomial $P_{p(\epsilon)}(k)=\left( \begin{array}{c} n \\ k \end{array} \right) p^k (1-p)^{n-k},$ which is asymptotically Poisson for $n>>1$, or $p<<1$.

\begin{figure}[t]
\includegraphics[width=0.43\textwidth]{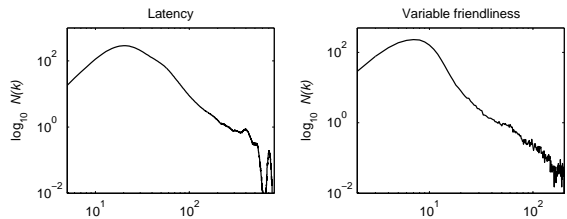} 
\caption{Either exponential latency (a) or exponential neighborhood size (b) can generate scale-free average distributions.}
\label{SFMN}
\end{figure}

A time-averaged scale-free network requires a substantially heavier tail than the basic MN model.  Thus motivated, and also considering that social connections, once formed, have certain persistence or memory, we model that some agents ``stay in touch,'' continuing to communicate for some period after they are no longer neighbors.  We formulate the following modification to MN, which we call Moving Network with Friends, or MNF:  To each agent we associate a random ``gregarious factor," $g_i=U(0,1)$.    As with MN, a new link is made between agents $i$ and $j$ whenever $|x_i-x_j|<\epsilon.$ However, once formed, we introduce latency as follows:  At each time step $T$ after $|x_i-x_j|>\epsilon$, we break the link $i \leftrightarrow j$ iff a uniform random $q=U(0,1)$ variable satisfies $q>F(g_j,g_i)=1-\sqrt{g_j g_j}$, where there is tremendous freedom in choosing $F$ depending upon the application, but we have chosen a specific form as matter of example here.  The exponential latency creates the power-law tail in the degree distribution, as shown in Fig.~\ref{SFMN}.  The early rise left of the maximum follows since our model still forms connections according to the  binomial distribution of MN, but now they are broken more slowly.  For large $k$, we find empirically that $P(k)\sim k^{-\alpha}$ with $\alpha\approx 2$.  An MNF, since it provides additional connectivity, has more robust synchronization properties than an MN network with the same neighborhood size, $r.$ 

It easy to formulate other MN-type models which produce a scale free structure, and we mention one more which we  find sufficiently applicable.  One can model that some nodes are ``friendlier" than others by defining the neighborhood of node $i$ to be of size $r_i,$ where $r_i$ need not be the same for every node.  An power-law distribution of $r_i$ would also generate a time-average scale-free network.  

{\bf Conclusions and Direction:}
In many real processes in which information propagation in ad hoc networks (such as disease spread, where the infective information may survive within an agent on the order of just weeks),  the recent network connections play a crucial role in the dynamic behavior of the system.  Thus we have been motivated to study time-evolving networks, which may more accurately describe the relevant dynamics. Our MN and MNF models provide a first attempt at developing such models, basing the network upon diffusing agents communicating within geographic neighborhoods and with established ``friends."  The numerical simulations in this paper show that global patterns (synchronization) are possible  in these models, {\it even when the network is spatially disconnected.}  We are developing a rigorous analysis of the moving average Laplacian to support  our empirical work, which well predicts stability criteria that can be related to previous network synchronization results.  Under the very general assumptions of ergodic network dynamics of the agents movements, we have proven the concept of an average degree distribution, and we have further shown that adding natural latency to network connectionism leads to the widely observed phenomenon of scale-free degree distribution, but now in a time-averaged sense, which is our new concept.  We expect these models to widely provide insight into relevant issues regarding swarming, flocking and other physical and technological ad hoc cooperative and emergent behavior, particularly if one expects the flock to act in some fashion that achieves a goal separate from the coordinated movement.  

EMB was supported by the National Science Foundation DMS-0071314.
Portions of this paper may be used by JDS as part of a doctoral dissertation.

{}



\begin{thebibliography}{}


\bibitem{NewmanReview} M. E. J. Newman,  SIAM Review 45, 167-256 (2003).

\bibitem{BarabasiReview}
R. Albert, A.-L. Barabasi, 
ÒStatistical mechanics of complex networksÓ, 
Rev. Mod. Phys. 74, 47 (2002)

\bibitem{Kuramoto} Y.~Kuramoto, {\sl Chemical Oscillations, Waves, and Turbulence,} Springer, Berlin, 1984.







\bibitem{erdosrenyi} P.~Erdos and A.~Reny\'{i}, Publ. Math. Inst. Hung. Acad. Sci., vol 5, 1960.







\bibitem{Watts} D.~J.~Watts, S.~Strogatz, Nature 393 (1998)







\bibitem{barabasi1} A.~L.~ Barab\'{a}si, R.~Albert, and H.~Jeong, Physica A, vol. 281, 2000. 







\bibitem{Lasota}

A.~Lasota, M.~Mackey, {\it Chaos, Fractals, and Noise, Second Edition} Springer-Verlag (New York, NY 1997).







\bibitem{Rossler} R\"{o}ssler OE. Phys Lett A, 1976; 57;397.







\bibitem{Barahona} M. Barahona and L. Pecora, Physical Review Letters, vol 89; 5, 2002.







\bibitem{Fink} K. Fink, G. Johnson, T. Carrol, D. Mar, L. Pecora,  Physical Review E, vol. 61;5, 2000.











\end{thebibliography}
\end{document}